\documentclass[aps,twocolumn, floats,preprintnumbers,showpacs,nofootinbib,prl]{revtex4}
\usepackage{graphicx}
\usepackage{url}
\usepackage{color}
\usepackage{enumerate}
\usepackage{amsmath}
\usepackage{amsfonts}
\usepackage{amssymb}

\newcommand{\bi}{\begin{itemize}}
\newcommand{\ei}{\end{itemize}}
\newcommand{\be}{\begin{equation}}
\newcommand{\ee}{\end{equation}}
\newcommand{\bea}{\begin{eqnarray}}
\newcommand{\eea}{\end{eqnarray}}

\def\bbaa{b\bar b \gamma\gamma}

\def\lhhh{\lambda^{hhh}}

\def\ifb{{\rm fb}^{-1}}
\def\iab{{\rm ab}^{-1}}

\begin{document}

\title{Higgs-Pair Production and Measurement \\of the Triscalar Coupling at LHC(8,14)}
\author{Vernon Barger$^{a}$, Lisa L.~Everett$^{a}$, C.~B.~Jackson$^{b}$, Gabe Shaughnessy$^{a}$\\
\vspace{2mm}
${}^{a}$Department of Physics, University of Wisconsin, Madison, WI 53706, USA
${}^{a}$Department of Physics, University of Texas at Arlington, Arlington, TX 76019, USA
}

\begin{abstract}
We simulate the measurement of the triscalar Higgs coupling at LHC(8,14) via pair production of $h$(125 GeV).  We find that the most promising $hh$ final state is $\bbaa$.  We account for deviations of the triscalar coupling from its SM value and study the effects of this coupling on the $hh$ cross-section and distributions with cut-based and multivariate methods.  Our fit to the $hh$ production matrix element at LHC(14) with 3 ${\rm ab}^{-1}$ yields a 40\% uncertainty on this coupling in the SM and a range of 25-80\% uncertainties for non-SM values.

\end{abstract}
\maketitle


{\it Introduction.}--- The long-awaited discovery of the massive particle ($h$) with Higgs-like characteristics at the LHC  \cite{Aad:2012tfa,Chatrchyan:2012ufa} heralds the beginning of a new era in particle physics.  The next experimental challenge is the measurement of the $h$-couplings to distinguish whether it is the Standard Model (SM) Higgs boson, or the lightest Higgs of the Minimal Supersymetric Standard Model (MSSM) or a general two Higgs doublet model (2HDM), or a state with an admixture of doublet and singlet components, or the lightest state of a more complex Higgs sector.  The answer to this question 
will have far-reaching implications about the existence and nature of any new physics at the TeV energy scale.  

In addition to the couplings of $h$ to gauge bosons, which are essential for the mass-generating mechanism, and the generation-dependent Yukawa couplings of $h$ to fermions, which are integral to $h$-production and its decays, the self-couplings of $h$ are of paramount interest since they directly connect to the underlying potential that results in spontaneous symmetry breaking.  In the SM, a single self-coupling parameter $\lambda$ completely specifies the potential, $V_{\rm SM} = -\mu^2 \phi^\dagger \phi + \lambda|\phi^\dagger \phi|^2$ and the Higgs mass is $m_h =  \sqrt\lambda v$, where $v$ is the vacuum expectation value (vev) of the Higgs field, which is determined by the Fermi coupling to be 246 GeV.  Based upon the Higgs mass measurement, $m_h = 125.5\pm 0.6$ GeV~\cite{ATLHMass}, the self-coupling value for the SM is $\lambda = 0.260\pm0.003$.  A precision measurement of the cubic coupling $\lhhh$ between three {\it physical} Higgs bosons is a priority of a linear $e^+e^-$collider, but this is more than a decade away.

\begin{figure}[t]
\centering
\includegraphics[scale=0.45]{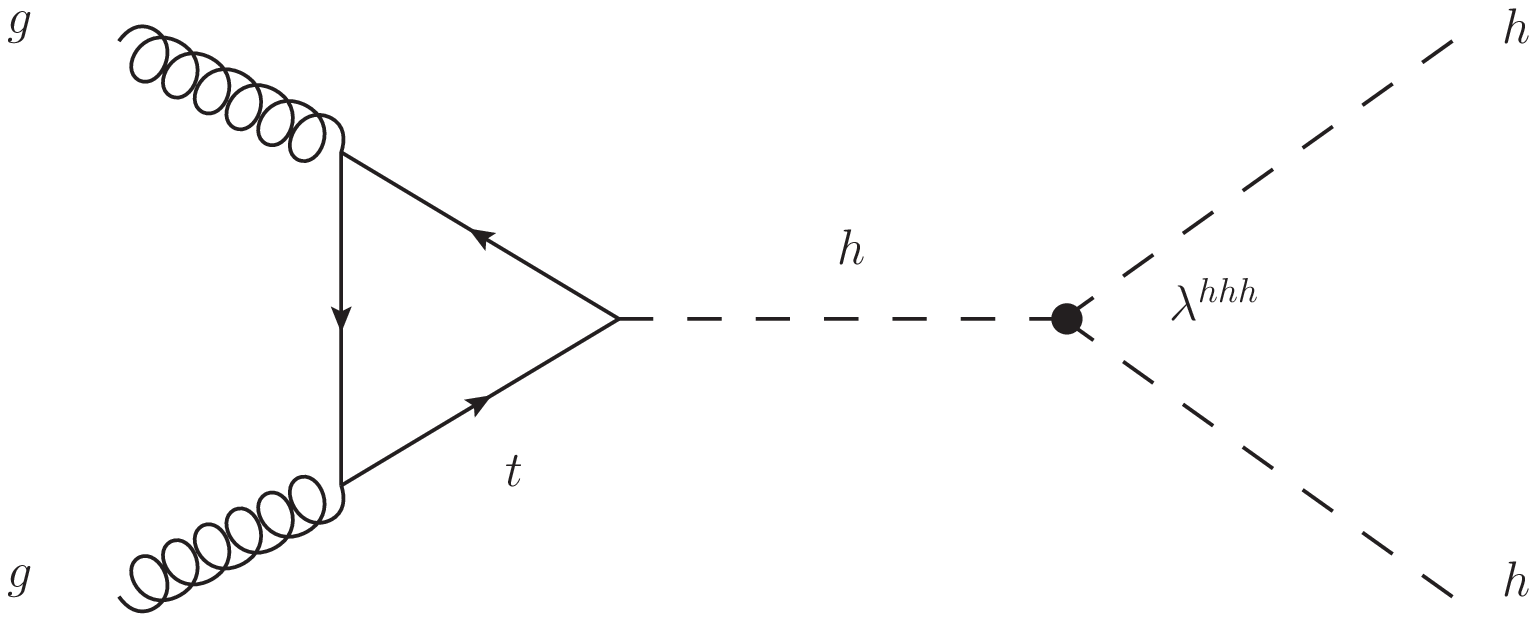}
\includegraphics[scale=0.45]{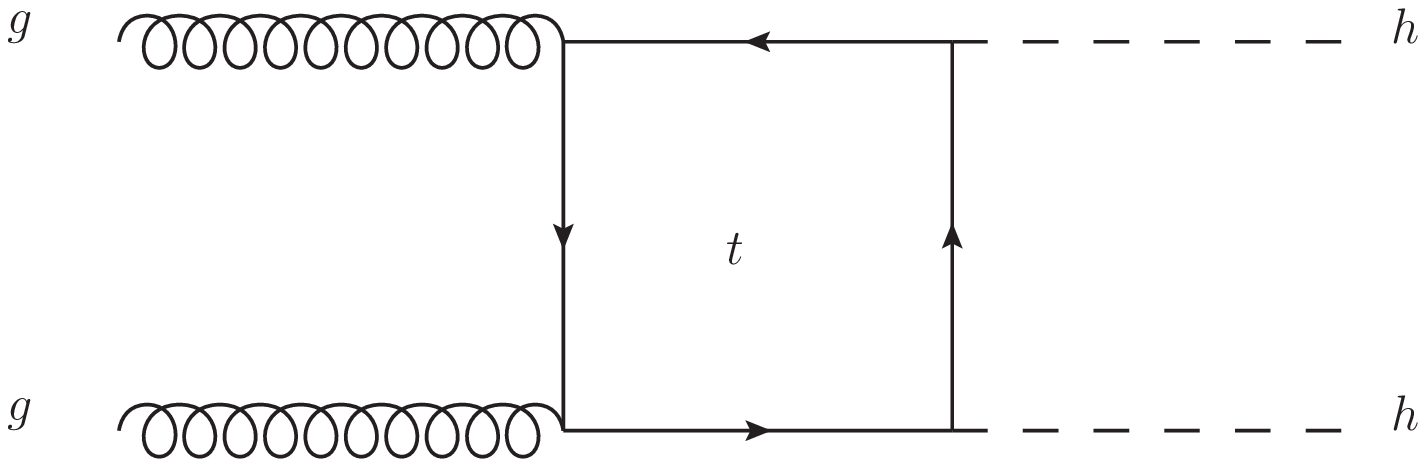}
\caption{Feynman diagrams which contribute to Higgs boson pair production via gluon fusion. }
\label{fig:feynman-diags}
\end{figure}

In a theory beyond the SM, there can be contributions to the effective potential from dimension six Higgs operators that are induced by integrating out heavy degrees of freedom, or from compositeness.  The Higgs mass and $\lambda$ then are independent parameters, and the
interactions of the Higgs with the electroweak gauge bosons are modified from their SM values.  An important goal is to measure  all of the Higgs self-couplings:  $hhh,  hhhh, hhWW$ and $hhZZ$.  The production of Higgs pairs at the LHC provides an important avenue to probe the first of these couplings, the triscalar coupling \cite{Djouadi:1999rca,Baur:2002qd,Moretti:2004wa,Arhrib:2009hc,Asakawa:2010xj,Grober:2010yv,Baglio:2012np,Dolan:2012rv,Contino:2012xk,Gupta:2013zza}, which we pursue in this letter.  The gluon-gluon fusion subprocesses of Fig.~\ref{fig:feynman-diags} are the dominant production diagrams \cite{Eboli:1987dy,Glover:1987nx,Dicus:1987ic,Plehn:1996wb}.  The interference of the two amplitudes is sensitive to the $hhh$ coupling and thereby provides a way to measure it.  We find that complete destructive interference of the real amplitudes occurs at $\lhhh \approx 2.45 \lhhh_{\rm SM}$.



\begin{figure}[h]
\begin{center}
\includegraphics[width=0.4\textwidth]{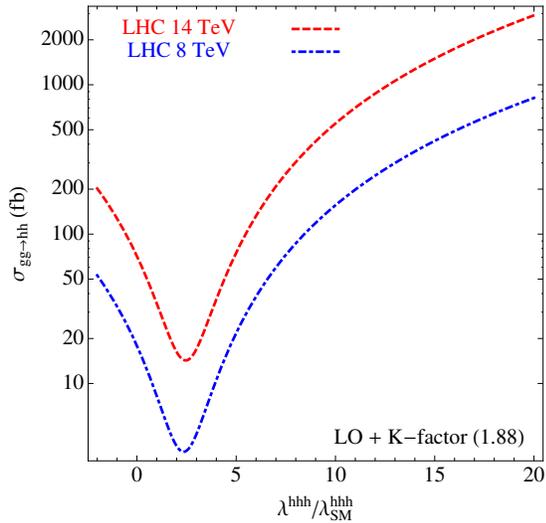}
\caption{Production cross section for $gg\to hh$ at the LHC with $\sqrt s = $ 8 TeV and 14 TeV.}
\label{fig:xs}
\end{center}
\end{figure}
\begin{figure}[t]
\begin{center}
\includegraphics[width=0.4\textwidth]{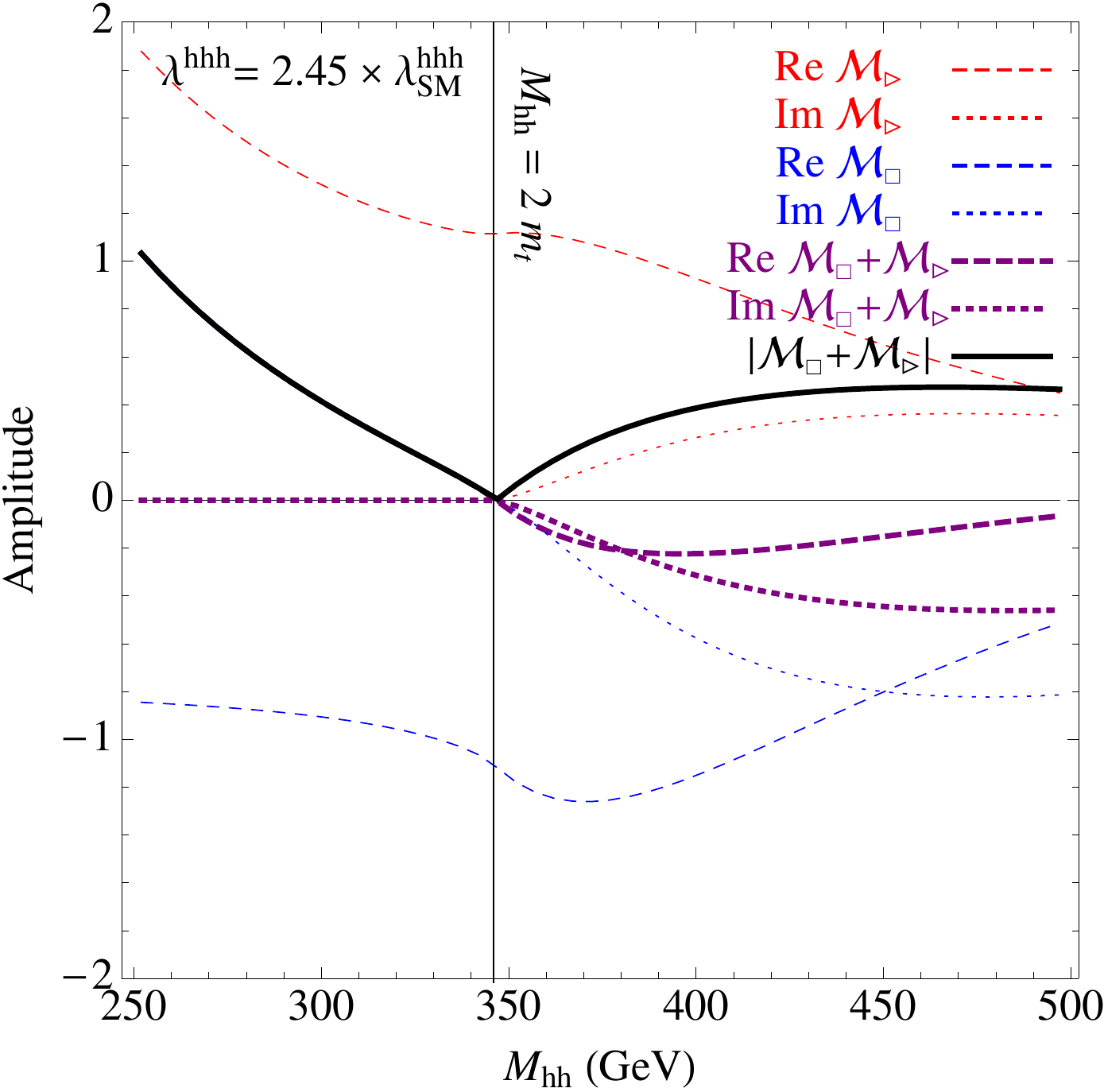}
\caption{Amplitude zero in $gg\to hh$ fusion versus $M_{hh}$ for $\lambda^{hhh}/\lambda^{hhh}_{\rm SM} =$ 2.45.  The SM value is $\lambda^{hhh}_{SM} = 192$ GeV.}
\label{fig:ampz}
\end{center}
\end{figure}

{\it Higgs pair-production cross section.}--- The leading order (LO) matrix elements of the $hh$ subprocesses in Fig.~\ref{fig:feynman-diags} are known \cite{Eboli:1987dy,Glover:1987nx,Dicus:1987ic,Plehn:1996wb}, up to the involved couplings.  We generate signal events by incorporating the loop amplitudes directly into MADGRAPH ~\cite{Alwall:2011uj}, and we include the NLO K-factor $=$1.88~\cite{Dawson:1998py,Dittmaier:2011ti,Branco:2011iw,deFlorian:2013uza}.  
The competition between the two diagrams in Fig.~\ref{fig:feynman-diags} strongly impacts the total cross section shown in Fig.~\ref{fig:xs} and the final state kinematic distributions, especially when the real parts of the two amplitudes cancel each other, as illustrated in Fig.~\ref{fig:ampz}. To account for possible new physics effects, we consider a broad range of $\lhhh$ values.  It can be shown that the high values of this range can be realized, for example, in general two Higgs doublet models wherein the additional doublet contributes to the triscalar coupling.

\begin{figure}[h]
\begin{center}
\includegraphics[width=0.4\textwidth]{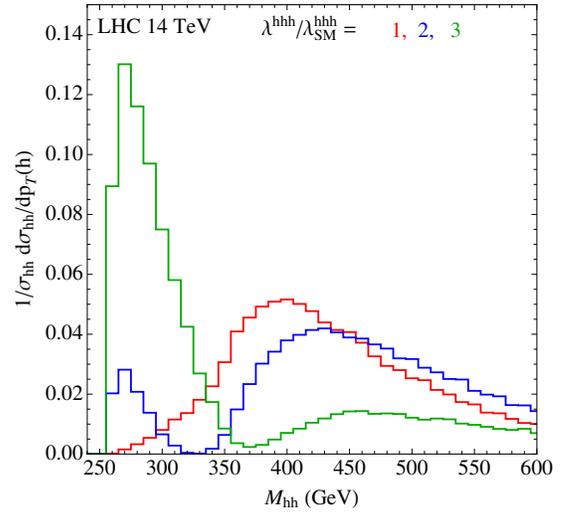}
\caption{The differential cross section versus $M_{hh}$ for $\lambda^{hhh}/\lambda^{hhh}_{\rm SM} =$ 1,2,3.} 
\label{fig:mhh}
\end{center}
\end{figure}
We calculate the $gg \to hh$ amplitudes for LHC center of mass energies of 8 TeV (we assume the relatively small data sample at 7 TeV is similar to the 8 TeV sample), for comparison with Run-1 data, and 14 TeV, for the upcoming high luminosity run.  The destructive interference occurs between the real parts of the triangle and box contributions.  For $1.1 \lesssim\lhhh \lesssim 2.45$, the cancellation of the real amplitude is exact at some value of $M_{hh}$.  The zero of the amplitude occurs at $M_{hh}$ near to $2m_t$; it is exactly at $2m_t$ for  $\lhhh \approx 2.45 \lhhh_{SM}$ as shown in Fig.~\ref{fig:ampz}.  Above the $t\bar{t}$ threshold, the amplitudes develop imaginary parts for which the cancellation does not occur.  Nonetheless, a  local minimum in the $M_{hh}$ distribution persists up to $\lhhh \approx 3.5\lhhh_{SM}$, and results in a rather low $M_{hh}$ dominated distribution, causing a large change in signal acceptance as we will see shortly.  The differential cross section, which is presented in Fig.~\ref{fig:mhh}, shows the persistence of the amplitude zero.  A related suppression is found to be present in the $p_T(h)$ distribution. 

For the Higgs decays, we consider the $\gamma\gamma$, $\tau\tau$, and $b\bar b$ modes, which are used in establishing the single higgs production signal \cite{Aad:2012tfa,Chatrchyan:2012ufa}.  Recently, there have been several studies of Higgs pair production using the $b\bar{b}\gamma\gamma$, $b\bar{b}\tau \tau$ and $b\bar{b}WW$ final states \cite{Baglio:2012np,Goertz:2013kp,Dolan:2012rv}. We do not study the $h$ to $W^+W^-$ decay as it contributes with low significance in $hh$ detection \cite{Baglio:2012np}. The signal of $hh\to\bbaa$ is robust with manageable background, so it is our primary interest.  The large backgrounds and combinatorics of the $hh \to b\bar b b \bar b$ final state render it unviable.  We also find the $b\bar b \tau_h \tau_h$ channel to be swamped by the reducible background of $b\bar b jj$ where both light flavored jets fake a hadronic $\tau$.  Although the jet to $\tau_h$ fake rate is only $1-3\%$, the total cross section of $b\bar b j j$ is at the $\mu b$ level. This insurmountable background was not considered in previous studies.   For this reason, we concentrate on the analysis of the $\bbaa$ channel and note that a more extensive study for the viability $\tau_h \tau_\ell$ and $\tau_\ell \tau_\ell$ is needed.

{\it Cut-based analysis for $hh \to \bbaa$.}---We simulate the pertinent backgrounds for the $\bbaa$ channel.  The irreducible backgrounds include the  production modes
\bea
pp&\to&\bbaa,\\
pp&\to& Z + h \to b\bar b +\gamma\gamma,
\eea
while the reducible backgrounds include
\bea
pp&\to& t\bar t + h ~\to~ b \ell^+ \nu ~\bar b \ell^- \bar \nu + \gamma \gamma \quad(\ell^\pm~ {\rm missed}),\\
pp&\to&b\bar b + j j ~\to~ b\bar b + \gamma \gamma\quad (j \to \gamma).
\eea
We adopt a photon tagging rate of 85\% and a jet to photon fake rate of $\epsilon_{j\to \gamma}=1.2 \times 10^{-4}$~\cite{Aad:2009wy}.  The additional reducible backgrounds from $jj\gamma\gamma$ and $c\bar c \gamma\gamma$  to be subdominant and hence are not  included in our analysis.
For $b$ jet tagging efficiencies, we assume a $b$-tag rate of 70\% for  $p_T{(b)}> 30~{\rm GeV}$ and $|\eta_{b}| < 2.4$,  as found in multivariate tagging estimates for the LHC luminosity upgrade~\cite{atlphyspub2013004}.  We adopt the $b$-mistag rate found in Ref.~\cite{Baer:2007ya}.  With pile-up, the background rejection rate is expected to worsen by up to 20\%~\cite{atlphyspub2013004}, which we take into account.  Finally, we model detector resolution by smearing the final state energy according to ${\delta E / E} = {a/\sqrt{E/{\rm GeV}}} \oplus b$, with $a=50\% (10\%)$ and $b=3\% (0.7\%)$ for jets (photons).

\begin{figure}[htbp]
\begin{center}
\includegraphics[width=0.4\textwidth]{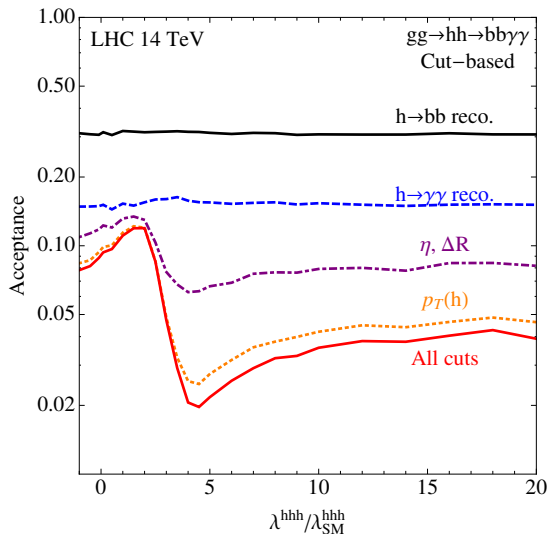}
\caption{Signal acceptance for the $\bbaa$ channel at various cut levels.  The reduced acceptance for large values of $\lambda^{hhh}$ correspond to a lower $p_T(h)$}
\label{fig:acc}
\end{center}
\end{figure}

We adopt the cuts in Ref.~\cite{Baglio:2012np} with:
\bea
p_T(b\bar b), \quad p_T(\gamma\gamma)&>&100~{\rm GeV},\label{eq:kincutpt}\\
M_{b\bar b \gamma\gamma} &>& 350~{\rm GeV}.
\label{eq:kincutmhh}
\eea
Our signal and background acceptances in the SM calculation agree with those of Ref.~\cite{Baglio:2012np}.  In Fig.~\ref{fig:acc} we show the acceptance at different cut stages versus the  $\lhhh$ coupling.  Note the prominent enhancement in the vicinity of $\lhhh=0 - 2$ and the suppression at $\lhhh>3$ which result from the $p_T(h)$ and $M_{hh}$ characteristics previously described.  For $\lhhh$ values that yield a significant enhancement in the Higgs-pair cross section, $\lhhh$ can be measured by the large event rate of the signal and the steep dependence of the signal cross section on $\lhhh$.  

The level of statistical significance, ${\cal S}$, is specified by
\be
{\cal S} = 2 \left(\sqrt{S+B}-\sqrt{B}\right),
\ee
where $S$ and $B$ are the number of signal and background events, respectively, which survive  the cuts.  This definition is less prone to downward fluctuations of the background~\cite{Bityukov:2000xx,Bartsch:2005xxa}.

{\it Multivariate analysis for $hh \to \bbaa$.}--- Here we consider an analysis based on the simultaneous multiple variables that in essence allows us to blend cuts together rather than perform specific hard cuts on kinematic distributions.  Such a multivariate discriminator can offer a sensitivity similar to that of the matrix-element or neural network methods~\cite{Abazov:2008kt}.  We form a discriminant ${\cal D}$ based on the following set of observables: ${\cal O}=\left\{ M_{\bbaa}, M_{b\bar b}, M_{\gamma\gamma}, p_T(b\bar b), p_T(\gamma\gamma), \Delta R_{b\bar b}, \Delta R_{\gamma\gamma}\right\}$.  The bayesian inspired discriminator is defined to be ${\cal D}$ =$S({\cal O} )/(S({\cal O}) + B({\cal O}))$, where $S$ and $B$ denote the signal and background differential cross sections~\cite{Bhat:2010zz,Gainer:2011aa}.  The discriminator is evaluated for a simulated event sample; it will be close to 1 for signal events and close to 0 for background events.  A cut may be placed on ${\cal D}$ to select a relatively high signal event sample.  
In practice, we apply a simplified version of the discriminator in which we ignore the correlations among the variables.  This allows a more efficient numerical estimation of the discriminator, defined as
\be
{\cal D}=\prod_{i=1}^{N} \delta_i {S({\cal O}_i)\over S({\cal O}_i) + B({\cal O}_i)},
\ee
where $N$ is the number of observables that in the multivariate discriminator and $\delta_i=\left\{0,1\right\}$ toggles the input of observable ${\cal O}_i$ into the discriminant.  Further optimization of the MVA would include correlations between observables.  We calculate the significance, ${\cal S}$, by placing a cut on the discriminator, ${\cal D}_{\rm cut}> 0.9$, and maximize ${\cal S}$ over all possible toggle states $\delta_i$.  We chose a fixed cut to be more conservative.  However, one can increase the singificance further by optimizing the discriminator cut; by optimizing ${\cal D}_{\rm cut}$, up to a 50\% additional improvement may be possible, allowing ${\cal S} > 5\sigma$ over the entire range of $\lhhh$ with 3 $\iab$ of integrated luminosity at the LHC14.

\begin{figure}[t]
\begin{center}
\includegraphics[width=0.39\textwidth]{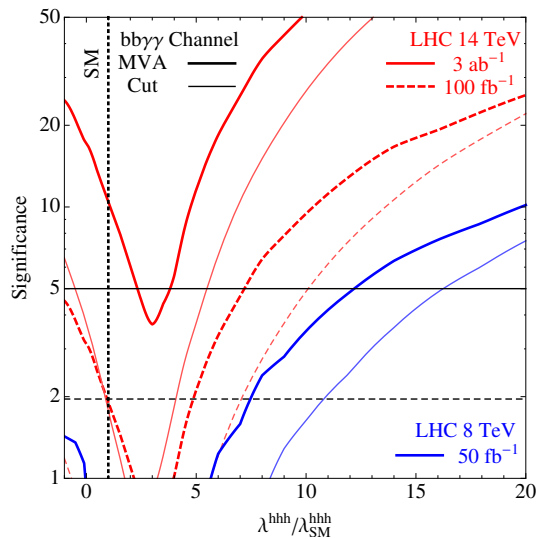}
\caption{Expected significance of the MVA (thick curves) and Cut-based analysis (thin curves) at LHC8 and LHC14.}
\label{fig:mva-vs-cut}
\end{center}
\end{figure}
In Fig.~\ref{fig:mva-vs-cut} we compare the MVA discovery reach with that of the cut-based analysis, at both LHC8 and LHC14.  It is apparent that the MVA gives superior performance. However, for large values of $\lhhh$, the increased reach of MVA is incremental, due to the dominance of the signal rate over the SM background.  This figure gives the luminosity needed for a $5\sigma$ discovery or for 95\% C.L. exclusion at LHC(7,8) with the accumulated 50 $\ifb$ luminosity, under the reasonable assumption of similar reach at 7 and 8 TeV.

The simulated coupling measurement is shown in Fig.~\ref{fig:coupfit}.  We take the matrix element for the LO process and, after unfolding the binned acceptance from simulated events with all cuts up to Eqs.~(\ref{eq:kincutpt})-(\ref{eq:kincutmhh}), fit the differential distribution, $d \sigma/d M_{hh}$.  We conservatively assume a 30\% and 100\% uncertainty in the signal and background cross section, respectively.  For the SM, we infer a coupling uncertainty of 40\%, which compares well with previous studies of $\delta \lhhh / \lhhh_{\rm SM} \approx 50\%$~\cite{Yao:2013ika}.  Otherwise, we find the achievable coupling uncertainty ranges from 25\% to 80\%, with the latter value due to the reduced cross section and acceptance in the region $\lhhh \sim 2-5$.  

\begin{figure}[htbp]
\begin{center}
\includegraphics[width=0.39\textwidth]{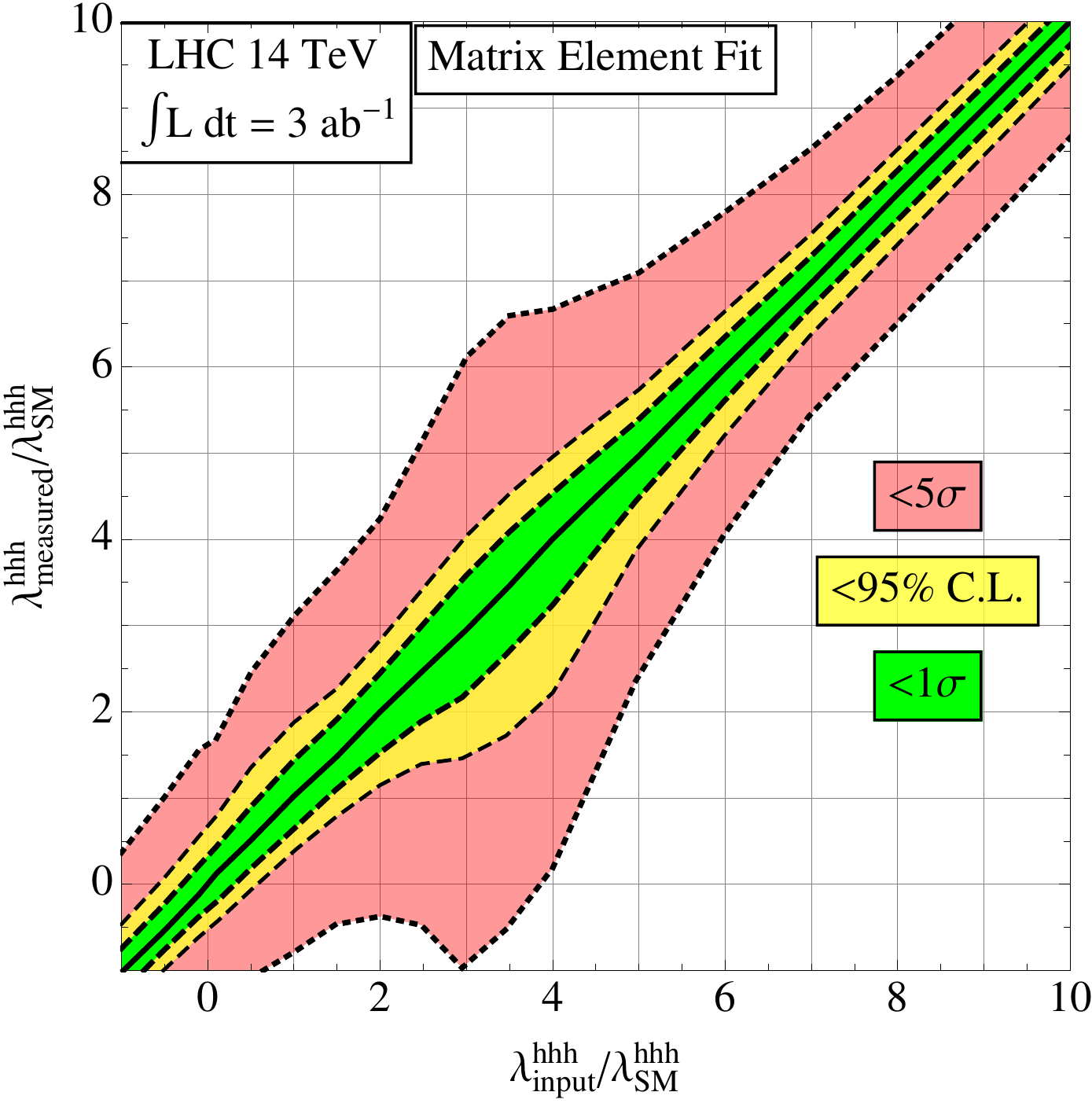}
\caption{Simulated coupling fit at LHC14 with 3 $\iab$ of integrated luminosity.}
\label{fig:coupfit}
\end{center}
\end{figure}

%


{\it Conclusions.}--- In this letter, we have investigated Higgs pair production at the LHC(8,14) as a probe of the Higgs triscalar coupling, $\lhhh$.
Our principal findings are as follows.  (i) 
The $b\bar b\gamma \gamma$ channel is the only promising channel; reducible backgrounds swamp the signals of other channels such as  $b\bar b \tau \tau$.
(ii) The minimum in the integrated cross section versus the triscalar coupling coincides with the minimum in the $M_{hh}$ distribution at $2 m_t$ for a $hhh$ coupling  $\lhhh\approx2.45 \lhhh_{\rm SM}$.
(iii) The amplitude of $gg \to hh$ has a zero in the $M_{hh}$ distribution for $1.1 \lesssim \lhhh / \lhhh_{\rm SM} \lesssim 2.45$.  A minimum occurs in the transverse momentum distribution $p_T(h)$ of each $h$ in $hh$ production. 
 (iv) Multivariate analysis gives a substantially better reach on $\lhhh$ over the cut-based analysis.
(v) LHC data at 7-8 TeV should probe large deviations of $\lhhh$ from the SM ($\lhhh/\lhhh_{\rm SM}\gtrsim 7.5$ at 95\% C.L.), while the 14 TeV data probes $\lhhh$ to 25-80\%.  At LHC14 with 3 ${\rm ab}^{-1}$,  $\lhhh_{\rm SM}$ can be determined within 40\% uncertainty. 

\section{Acknowledgments}
VB, LLE and GS are  supported by the U. S. Department of Energy under the contract DE-FG-02-95ER40896.  GS thanks J.~Gainer for useful discussions.

\appendix

  \end{document}